\documentclass[aps,prl,reprint]{revtex4-1}
\usepackage{tikz}
\usepackage{graphicx}
\usepackage{amsmath}
\usepackage{amssymb}

\begin{document}

\title{Recurrent Neutrino Emission from Supermassive Black Hole Mergers}
\author{O.~de Bruijn$^{1,2}$, I.~Bartos$^{3}$, P.L.~Biermann$^{4,5,6,7}$, J.~Becker Tjus$^{1,2}$}
\affiliation{
$^1$ Theoretical Physics IV: Plasma-Astroparticle Physics, Faculty for Physics \& Astronomy, Ruhr-Universit\"at Bochum, 44780 Bochum, Germany\\
$^2$ Ruhr Astroparticle And Plasma Physics Center (RAPP Center)\\
$^3$ Department of Physics, University of Florida, PO Box 118440, Gainesville, FL 32611-8440, USA\\
$^4$ MPI for Radioastronomy, 53121 Bonn, Germany\\
$^5$ Department of Physics, Karlsruhe Institut für Technologie, 76344 Karlsruhe, Germany\\
$^6$ Department of Physics \& Astronomy, University of Alabama, Tuscaloosa, AL 35487, USA\\
$^7$ Department of Physics \& Astronomy, University of Bonn, 53115 Bonn, Germany
}

\begin{abstract}
The recent detection of possible neutrino emission from the blazar TXS 0506+056 was the first high-energy neutrino associated with an astrophysical source, making this special type of active galaxies promising neutrino emitters. The fact that two distinct episodes of neutrino emission were detected with a separation of around 3 years suggests that emission could be periodic. Periodic emission is expected from supermassive binary black hole systems due to jet precession close to the binary's merger. Here we show that if TXS 0506+056 is a binary source then the next neutrino flare could occur before the end of 2021. We derive the binary properties that would lead to the detection of gravitational waves from this system by LISA. Our results for the first time quantify the time scale of these correlations for the example of TXS 0506+056, providing clear predictions for both the neutrino and gravitational-wave signatures of such sources.
\end{abstract}

\maketitle

\section{Introduction\label{intro:sec}}
With the first detections of astrophysical high-energy neutrinos \citep{icecube2013,icecube2014} and gravitational waves \citep{gw_bhmerger2016}, the era of multi-messenger astronomy has begun, enabling the exploration of the Universe in a whole new manner. We have learned from gravitational-wave observations that there exists a significant population of stellar-mass binary black holes with properties we only begin to understand \citep{2019PhRvX...9c1040A,PhysRevD.100.064064,2019PhRvD.100f4064A,2020arXiv200408342T}. Neutron star mergers can now be investigated using a combination of gravitational waves and the broad photon spectrum from radio up to GeV gamma-rays, in the future possibly adding TeV gamma-rays and neutrinos \citep{gw_nstars_2017,2013CQGra..30l3001B,KohtaBartos,2017ApJ...850L..35A,2019ApJ...870..134A}. We have learned from neutrino observations that the high-energy events are of extragalactic origin given the lack of clustering of events in the Galactic plane \citep{numu_signal2016,icecube_cascades2019}. The detection of a possible correlation of a gamma-ray flare from the blazar TXS 0506+056 with a high-energy neutrino in September 2017 is a first hint that active galaxies may produce a significant fraction of the observed flux of cosmic high-energy neutrino \citep{icecube_txs_fermi2018}. A dedicated search for further neutrino flares from TXS 0506+056 in the past revealed a $3.5\sigma$ evidence for a long-duration ($110^{+35}_{-24}$~days) flare of TeV neutrinos around 2.8 years prior to the 2017 measurement \citep{icecube_txs2018}. No gamma-ray flare occurred during that time, which makes the modeling of the multi-messenger data challenging \citep{reimer2019}, although not impossible \citep{halzen2019}. Several ideas of how to produce these multi-messenger signatures have been presented, see e.g.\ \citep{murase_padovani2019,cao2019,xue2019}.

Transient blazar flares may be produced in the wake of galactic mergers when two supermassive black holes (SMBHs) inspiral towards each other. The black holes on close orbits reorient their spin in this inspiral phase. Spin precession due to relativistic effects in turn can result in precessing relativistic outflows that periodically change the direction of high-energy radiation \citep{gergely_biermann2009}. This scenario leads to the prediction of a population of blazars that are currently in such a state. The first potential hints of such signatures have been identified by Kun et al.\ \citep{kun2017}. Recent observations of periodicity at radio wavelengths also point toward a precessing jet scenario \citep{britzen2019}. 

In this {\it Letter}, we examine the observational consequences of transient blazar flares could arise  from jet precession in supermassive binary black holes (SMBBHs) close to merger. In particular, we examine the possible time structure of high-energy neutrino emission from blazars due to jet precession, focusing on TXS 0506+056. Considering this scenario we predict the time of the next neutrino flare from  TXS 0506+056 and the expected time of the corresponding SMBBH merger. We discuss the possibility of detecting this merger through gravitational waves using the Laser-interferometer Space Antenna (LISA) in the next decade: LISA will be the first gravitational wave detector in space, consisting of three spacecrafts that are positioned in an equilateral triangle. The gravitational frequency range probed by LISA is in the $0.1$~mHz to $1$~Hz range and well-suited for the detection of supermassive black hole mergers. The launch of LISA is scheduled in the early 2030s. Here, we discuss the parameter range for which gravitational waves from TXS 0506+056 occur in the time frame of the lifetime of LISA.
\section{Jet precession at supermassive binary black holes\label{intro:sec}}
Many accreting supermassive black holes are observed to drive relativistic jets. The rotation angle of the accretion disks around these black holes, and consequently the direction of the jets are thought to be aligned with the spin axis of the black holes \citep{Rees1978}. When two supermassive black holes inspiral towards each other, their spin precession will also change the orientation of the jet periodically.

\begin{figure}
 \includegraphics[width=\linewidth]{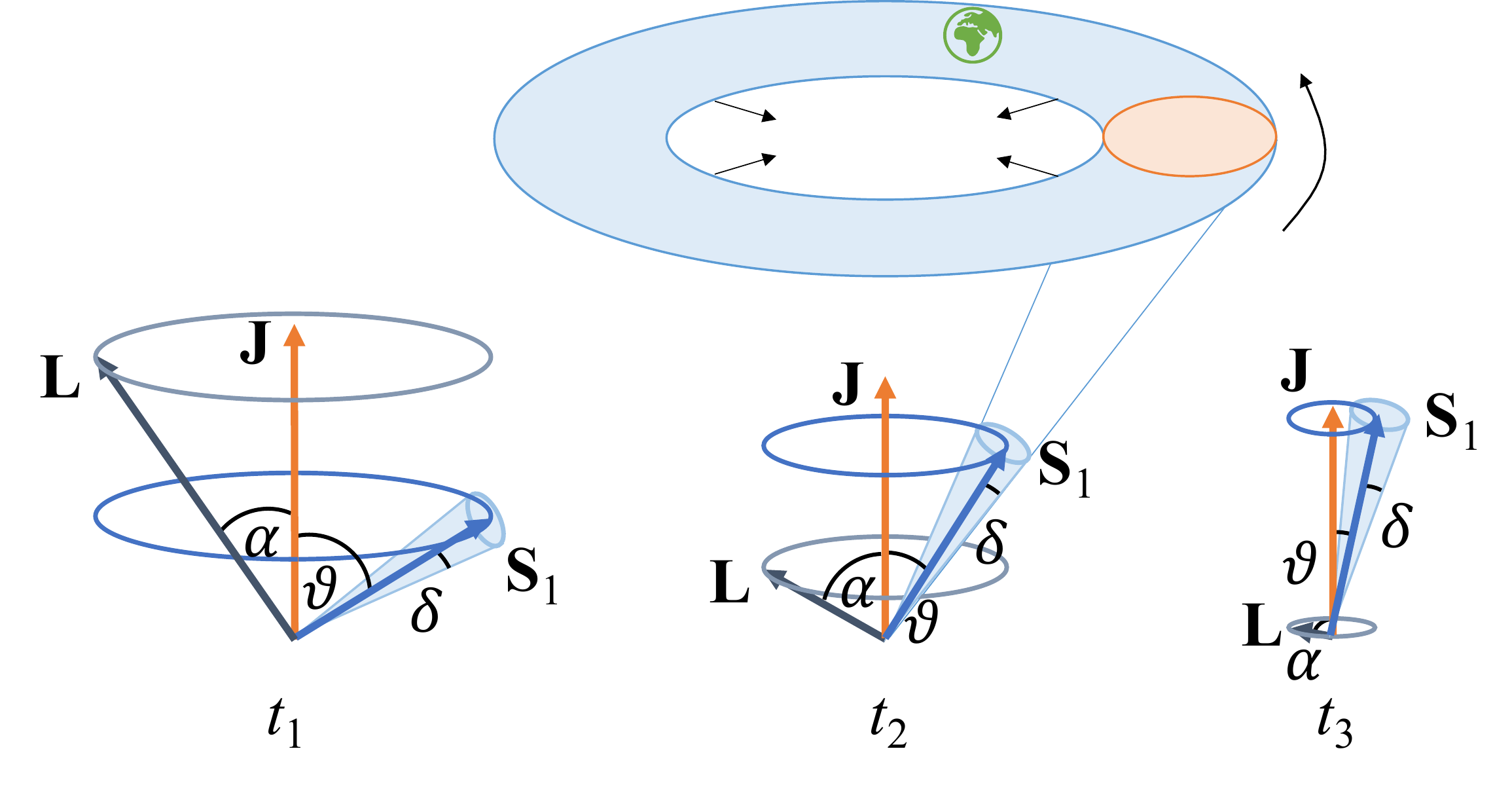}
 \caption{Schematic representation of the jet precession during the rearrangement of the jet axis. The blue cone represents the strong jet with the half opening angle $\delta$ around the spin axis $\mathbf{S}_1$. The weaker jet $\mathbf{S}_2$ is not shown. The jet is precessing around the total momentum axis $\mathbf{J}$ and the spin $\mathbf{S}_1$ aligns with $\mathbf{J}$ with time. In the scenario presented here, the line of sight is oriented in a way that the jet points toward Earth periodically.  \label{precession:fig}
  }
 \end{figure}

We model the inspiral dynamics of a SMBH binary close to merger using the post-Newtonian (PN) approximation up to 2.5 PN order (for a review see \citep{Blanchet2014}). The magnitude of the spins of the two black holes ($i=1,\,2)$ are defined as:
\begin{equation}
	S_i \approx \frac{Gm_i^2}{c}\chi_i,\quad i=1,2 \label{eq:Spins}
\end{equation}
with the dimensionless spin-parameter $ \chi_i = V_i/c\in [0,1] $ and the rotational velocity  $V_i$. We assume, that the rotation velocity is maximal, i.e.\ $V_i\approx c$, leading to $\chi \approx 1$. Comparing $S_1$ and $S_2$ leads to:
\begin{equation}
	\frac{S_2}{S_1} \approx \left(\frac{m_2}{m_1}\right)^2 = q^2.
\end{equation}
Here, we can neglect the spin of the lighter black hole as $S_2\ll S_1$ holds for most binaries, in particular concerning the  range of mass ratios $0.033\lesssim q\lesssim 0.33$ that has been identified as most common in  \citep{Caramante2010,gergely_biermann2009}. Due to the spin-orbit interaction, which is a 1.5 PN order effect, the axes of the angular momentum $\mathbf{L}$ and the spin $\mathbf{S}_1$ start precessing around the axis of the total angular momentum $\mathbf{J}$ \citep{gergely_biermann2009}.  This is the dominant effect and just slightly modified by additional interactions, i.e. spin-spin, mass quadrupolar, magnetic dipolar, self-spin and higher order spin-orbit effects, which we can neglect as shown in \cite{gergely_biermann2009}. The precession process starts when the timescale of gravitational radiation becomes comparable to the timescale of dynamical friction, see e.g.\ \citep{zier2001,zier2002,bar-or2014,bar-or2016}. As the black hole's distance gradually decreases, $\mathbf{S}_1$ eventually aligns with $\mathbf{J}$ (see Fig. \ref{precession:fig}).

The timescale of gravitational radiation, defined here as the remaining time till merger \mbox{$\Delta T_{\rm GW}=t_{\rm merger}-t$}, is used as a running variable as it is defined by the angular momentum $L$ and its change rate $\dot{L}$ at a given time $t$:
\begin{equation}
    \Delta T_{\mathrm{GW}}=-\frac{L}{\dot{L}}\approx
    \frac{5 G M}{32 c^{3}} \varepsilon^{-4} \eta^{-1}\,,
\label{delta_tgw:eq}
\end{equation}
with the total mass of the binary $M$, the post-Newtonian parameter $\varepsilon \approx v/c$ and the symmetric mass ratio $\eta = m_{1} m_{2}/(m_{1}+m_{2})^{2}$. 

Angular momentum loss due to gravitational radiation becomes dominant once $\Delta T_{\mathrm{GW}}$ becomes comparable to the characteristic time of dynamical friction due to the dense stellar cusp in the galactic center. We can use this to set a lower bound on $\varepsilon$ in the precessing regime  \citep{gergely_biermann2009}:
\begin{equation}\label{eq:Inspiral_criterion}
\varepsilon> \varepsilon^{\star}=\left[\frac{45}{64}(1+q)\right]^{2 / 11}\left(\frac{G M}{c^{2} r_{\text {distr }}}\right)^{6 / 11}\approx 10^{-3}.
\end{equation}
Here, we use $r_{\rm distr}=5$\,pc for the core radius of the central, compact stellar distribution \citep{zier2001}.

Our Post-Newtonian approximations of order 2.5 are sufficiently accurate for $\epsilon\lesssim 0.1$, corresponding to $\Delta T_{\rm GW}\gtrsim 0.5-2.4$\,yr before the merger. This regime is sufficient for the purposes of our calculations below, in particular in the case of TXS 0506+056.

Once the system loses enough orbital angular momentum so that $L<S_1$, the angular velocity of the precession $\Omega_P$ can be expressed as \citep{gergely_biermann2009}:
\begin{equation}\label{eq:Omega_approx}
    \Omega_{\mathrm{P}}(\varepsilon)\approx\frac{2 c^{3}}{G M} \varepsilon^{3}\,.
\end{equation}
We use this to obtain the angle $\phi$ of $S_1$ as a function of time until merger:
\begin{equation}
\label{eq:phi}
    \phi(\Delta T_{\rm GW})=
    -8 \left(\frac{5c}{32\eta\,G^{1/3}M^{1/3}}\right)^{3/4}\Delta T_{\mathrm{GW}}^{1/4}+\phi_0\,.
\end{equation}
Here, $\phi_0$ the integration constant, fixed for the specific problem.
 \section{Predictions for neutrino emission and merger time for TXS 0506+056 \label{predictions:sec}}
The blazar TXS 0506+056 produced a detectable neutrino signature both in 2014/2015 \citep{icecube_txs2018} and then in September 2017 \citep{icecube_txs_fermi2018}. 
We now examine the possibility that TXS 0506+056 is a supermassive black hole binary system in which the two observed neutrino emission episodes are due to the precession of the relativistic outflow from the black holes. We can use observations in the context of this scenario to make predictions on upcoming flares, and even the time of the binary merger.

\begin{figure}
 \includegraphics[width=\linewidth]{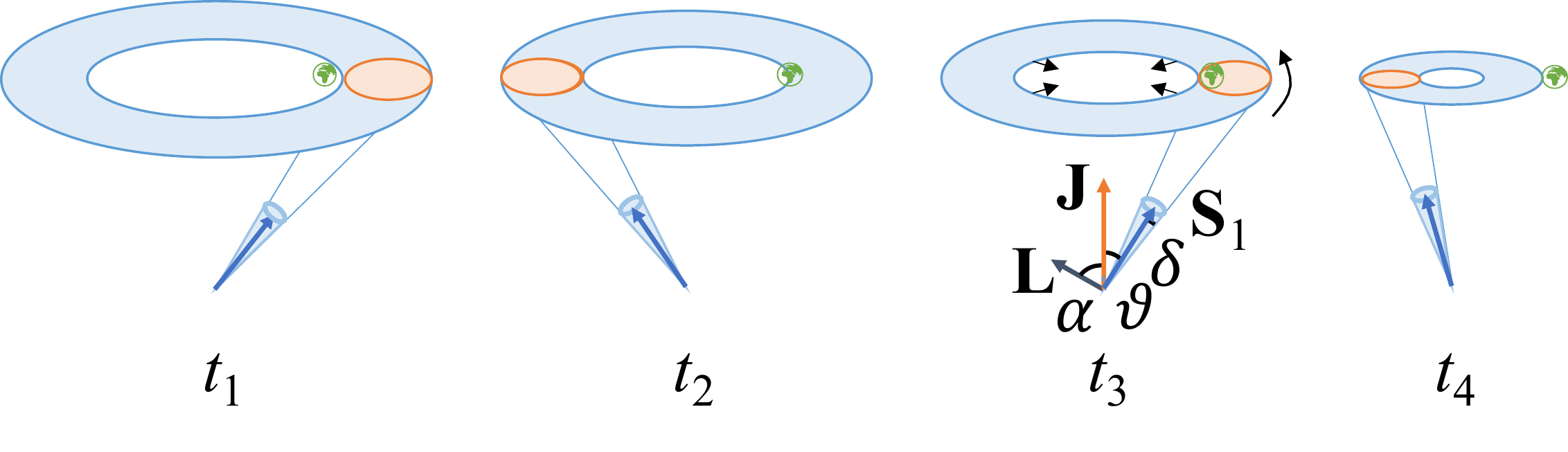}
 \caption{Representation of the signal-structure an AGN jet that recurrently points at Earth. Four time steps are shown, indicating that the periodicity of the events seen from Earth decrease with time as the angle between the angular momentum $\mathbf{J}$ and the dominant spin $\mathbf{S_1}$, i.e.\  $\mathbf{L}\cdot \mathbf{S_1}=L\,S_1\cos\theta$, decreases gradually. The first time step shows the time in which the periodic flares can be detected for the first time, as the emission cone starts to cross Earth. The fourth time step shows the last detection of a flare, i.e.\ the time at which the jet does not point at Earth anymore. The time difference betwee $t_4-t1_1$ defines the time interval in which a blazar is pointing directly at Earth and recurrent neutrino emission can be observed. 
 \label{earth:fig}
  }
 \end{figure}

We model TXS 0506+056 as a  binary in which the jet of the more massive black hole is precessing around the total angular momentum $\mathbf{J}$. As indicated in Fig.\ \ref{earth:fig}, we expect a neutrino flux every time the jet points towards Earth. We use the time difference of the two potential neutrino flares to determine the current periodicity of the object as $2.78\pm 0.15$~years. To determine this periodicity, we assume that the duration of the two flares is generally approximately the same  and we base the length of the episodes on the one of the first flare: this one has a Gaussian width of $55$~days, as indicated by the Gaussian analysis of the flare \citep{icecube_txs2018}. For the second flare, we assume that the one detected event could have arrived any time during the duration of the period in which the jet points towards the Earth. We then assume that the duration of the 2017-flare is approximately the same as the 2014/2015 flare. This gives us an uncertainty of the duration of the period of $\pm 110/2$~days~$=\pm 0.15$~yrs, as the neutrino could have arrived earlier or later, with no specific preference.

These observables - i.e.\ the times and length of the flares - are used to fix the period of the path $\phi(t)$ and thus link the time variable and $\phi$ to the model with the two TXS outbursts. This enables us to predict the next neutrino signals and to calculate the merger time of the SMBBH. The other angles $\alpha$ and $\theta$ as indicated in Figures \ref{precession:fig} and \ref{earth:fig} are fixed, as the condition $\alpha+\theta=$~constant prevails (see \citep{gergely_biermann2009} for a discussion). Thus, only $\theta$ evolves independently and determines for how long a signature can be detected. For all possible parameter settings, we find that the jet periodically points back at Earth for the next $\sim 15-120$~years.  

Parameters that enter the calculation of $\phi(t)$ are the total mass of the system which we take to be $M=3\cdot 10^{8}\,M_ {\odot}$ \citep{Padovani2019} and the mass ratio of the two black holes, which is varied between $0.033<q<0.33$ as discussed above. 

The analytical description of the system works properly for times at which the angular momentum $L$ is smaller than the dominant Spin $S_1$, $t\gtrsim t({L<S_1})$. For values $q<0.2$, this is granted for all times. For larger values, $q\gtrsim0.2$, we need a description at smaller times and achieve this via extrapolation.

\begin{figure}
 \includegraphics[width=\linewidth]{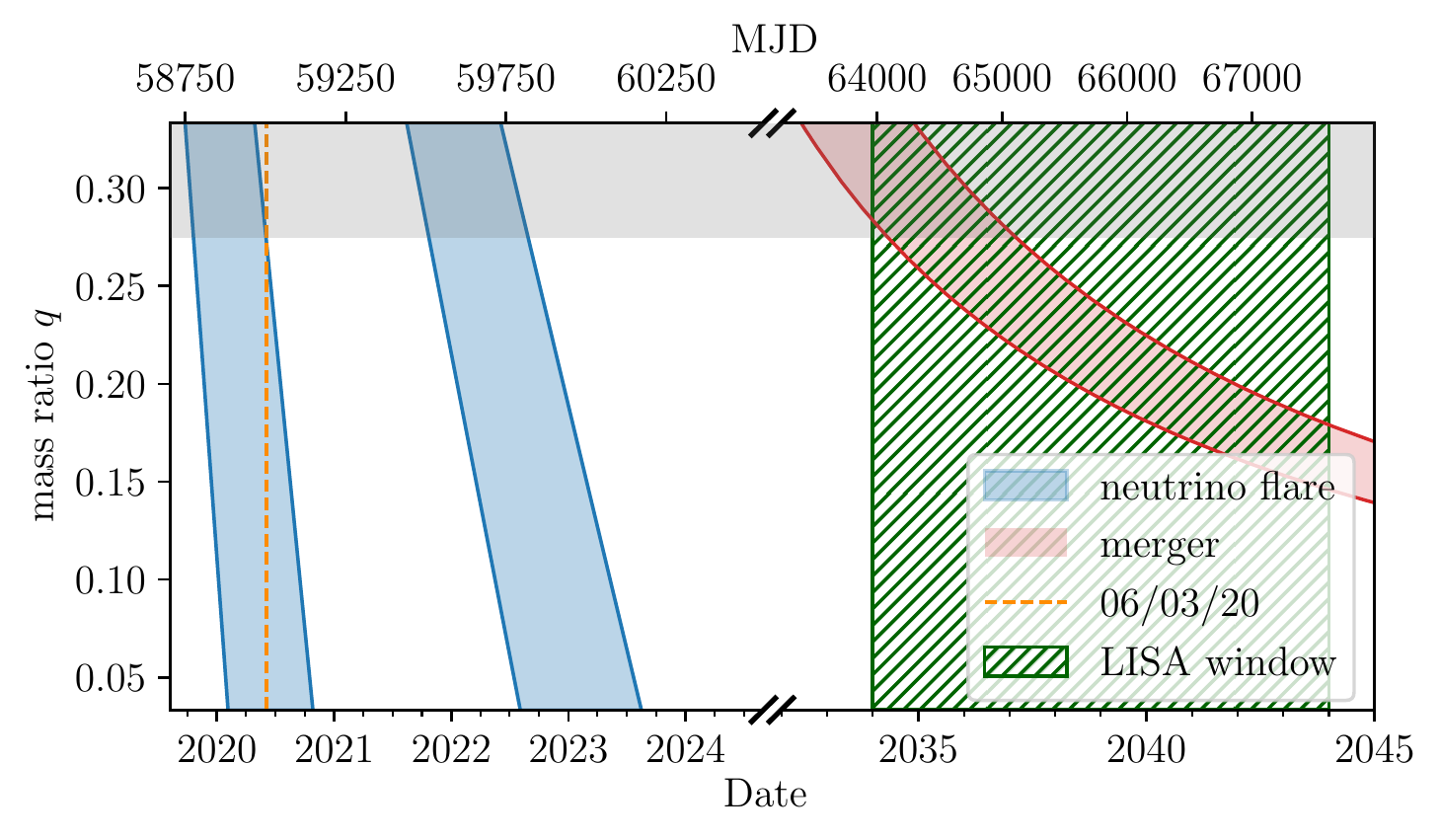}
 \caption{Prediction of the times of the next two neutrino flares (blue bands) and the time of the gravitational wave signature (red band) of the merger for the source TXS 0506+056. The dashed, orange line shows the date of \today. Based on the fact that no neutrino flare has been observed up to \today, the exclusion region of the parameter space is colored in grey. The green shaded area shows the expected up-time of LISA.
 \label{nu_gw_prediction:fig}
  }
 \end{figure}

The result of our calculation is shown in Fig.\ \ref{nu_gw_prediction:fig}. The blue shaded areas represent the predictions for the next two expected future periods in which the jet will point towards the Earth and when we would expect a new neutrino flare. The uncertainty band is shown in light-blue. As can be seen in the figure, the new time depends on the binary mass ratio
and the next flare should arrive before the end of October 2020. A dedicated unblinding of neutrino data for the time period 2019 - 2020 for the TXS 0506+056 could thus reveal another flare. The exact timing of such an event would enable us to specify the mass ratio of the two black holes and with high precision predict the flare after the next one. Due to possible intrinsic variations in the jet, however, it is not clear how strong the signal will be.
Figure \ref{nu_gw_prediction:fig} also shows the prediction for the binary merger time (red shaded area), and the expected operation time frame for LISA. For large mass ratios ($q\gtrsim 0.15$), if the origin of the two neutrino emission episodes is binary jet precession, LISA will be able to detect gravitational waves from TXS 0506+056 and thus enable the detection of neutrinos and gravitational waves from the same source. 

\section{Future population study of SMBH merger neutrino and GW events? \label{population:sec}}
In the scenario presented above, the TXS source is a candidate for a merger event of SMBBHs during the lifetime of LISA. But is this a singular event, or can we expect more sources that are neutrino emitters today and that could be detected as mergers in one or two decades? 

To present an order of magnitude estimate, we compute the expected merger rate $R_{\rm merger}\sim N_{\rm SMBHHs}/t_{\rm merger}^{\rm av}$, with $N_{\rm SMBBH}$ as the number of SMBBHs with their jet pointing toward Earth and $t_{\rm merger}^{\rm av}$ as the average timescale for the inspiral phase. The latter can be inferred from radio-interferrometric data as described in \citep{gergely_biermann2009}, with an estimate of $t_{\rm merger}^{\rm av}\sim 5\cdot 10^{6}$~yr.

In order to estimate the number of observable SMBBHs, we start with the sample of blazars at GHz frequencies (e.g. \citep{Eckart1986} and \citep{Eckart1987}). For blazars as sources with their relativistic jets pointing toward Earth, synchrotron emission is quite intense at these frequencies and thus a good tracer for these sources, 
 see e.g.\ \citep{Munoz2003} and references therein. About 320 sources down to 1 Jy at 5 GHz over all sky \citep{1JyCatalogue} are known. In the southern sky, these numbers correspond to a lower flux density limit (\citep{ParkesCatalogue}, \citep{Otrupec1991}). 

In order to investigate the full statistics, we need to consider the redshift evolution of the sources, following \citep{Biermann2014,Becker2008}. We use the cosmic activity rate as measured by \citep{Planck2014} as a redshift distribution of the sources, which indicates that the dominant contribution from blazars comes from the redshift range 1 to 2, with a broad peak up to $z=3$ contributing to the $\nu-$GW connection. We further apply the luminosity function presented in \cite{Caramante2010}.

Including all sources with a flux down to  $10^{-3}$ Jy at 5 GHz and higher could increase the sample size from about $10^{2}$ to about $10^{6.5\pm 2}$.  If the redshift evolution is weaker than the general one seen in Planck data \citep{Planck2014}, this would push the redshift range of the dominant contribution to redshifts $z<1$. This way, the number of sources would decrease by an order of magnitude, while the sources would, on the other hand, contribute with a stronger signal per source as they are closer. 

Concering the question on how many of these sources really are neutrino emitters, 
\citep{halzen20192} summarizes that only a fraction of the population of known GeV gamma-ray blazars is sufficient to explain the neutrino background. If every source would be as strong as TXS, about 5\% of the sources would be enough. As the other neutrino sources are not identified yet, they are expected to be weaker and thus, the true fraction will lie significantly above 5\%. Stacking-searches with IceCube show that only $27\%$ of the GeV blazars can contribute to the IceCube diffuse neutrino flux, so we summarize that it is expected that the contribution is somewhere between $5\% - 27\%$.

With the discussion above, we conclude that the number of active sources that we can expect in the Universe to be observable right now as precessing SMBBH jets in electromagnetic waves and neutrinos should be on the order of $10^{5}<N_{\rm SMBBH}<10^{8}$. With the average inspiral time of $t_{\rm merger}^{\rm av}$, the merger rate can therefore be estimated to $0.03$~yr$^{-1}<R_{\rm merger}<30$~yr$^{-1}$.

Within these uncertainties, our conclusions are as follows:
\begin{itemize}
    \item it is a plausible scenario to expect several merger events with GW signatures that also show electromagnetic and neutrino emission on time scales that suite the LISA time window.
    \item those sources that are close to the merger should reveal a time periodicity in the electromagnetic and neutrino spectra already now, $15-25$~years before the actual merger. To identify such a periodic behavior is only possible if intrinsic variations of the sources are not too strong.
\end{itemize} 

\section{Summary and outlook\label{discussion:sec}}
Explaining the two neutrino flares of the blazar TXS 0506+056 in the scenario of an AGN jet that is precessing prior to the merger of two supermassive black holes leads to three key predictions:
\begin{enumerate}
    \item A new neutrino flare is expected any time between late 2019 and the end of October 2020.  A blind analysis of the IceCube data or other detectors that search for neutrinos (see e.g.\ \citep{auger2019}) with data from the direction of TXS 0506+056 could therefore reveal a flare if the signal is strong enough to be detectable. 
    \item If the mass ratio of the binary system of TXS 0506+056 is $q\gtrsim 0.15$, the merger will occur at a time when it can be detected by LISA. For mass ratios $q\lesssim 0.15$, merging will happen at later times, up to $120$~years from now.
    \item We estimate the merger rate of SMBBHs that drive precessing, relativistic outflows to be $\sim1$\,yr$^{-1}$,  albeit with large uncertainties. Therefore, the explanation of the neutrino detection from TXS as a precursor event for a SMBBH merger is reasonable in the light of these numbers. 
    The example of TXS shows that the periodicity of those sources that will produce gravitational-wave emission in $\sim 15-25$~years, is around a few years. Thus, in order to identify more neutrino sources that might also be observable through gravitational waves later, such periodicity time scales will be interesting to search for with dedicated analyses using IceCube and future telescopes like Km3NET and IceCube-Gen2.
\end{enumerate}
The above predictions can open a new window to precision multi-messenger astrophysics if it is possible to identify sources that show a flaring behavior as discussed here. These observations can have the power to finally provide fundamental parameters of the SMBHs as the total mass, mass ratio and precession angle. The next years will show if the scenario of a precessing jet is viable for the system TXS 0506+056 and if other sources with jet precession can be detected even in neutrinos.

\begin{acknowledgments}
The authors would like to thank Francis Halzen, Emma Kun for valuable discussions on the topic and Roger Clay and Elisa Lohfink for commenting on the manuscript. IB is grateful for the support of the National Science Foundation under award number PHY-1911796, the Alfred P. Sloan Foundation, and the University of Florida Excellence Award. 
\end{acknowledgments}

%

\end{document}